\documentclass[aps,pra,showpacs,twocolumn]{revtex4}
\usepackage{graphicx}
\usepackage{amsmath, amsthm, amssymb}
\allowdisplaybreaks[4]
\newcommand{\vect}[1] {\mathbf{#1}}
\newcommand{\dif} {\mathrm{d}}
\newcommand{\I}{\mathrm{i}}
\newcommand{\up} {\uparrow}
\newcommand{\down} {\downarrow}

\newcommand{\brac}[1] {\langle #1\rvert}
\newcommand{\ket}[1] {\lvert #1\rangle}
\newcommand{\bracket}[2] {\langle #1 \rvert #2\rangle}

\newcommand{\erf} {\mathrm{erf}}
\begin{document}

\title{Large momentum part of fermions with large scattering length}
\date{\today}
\author{Shina Tan}
\affiliation{James Franck Institute and Department of Physics,
  University of Chicago, Chicago, Illinois 60637}

\begin{abstract}
It is well known that the momentum distribution of the two-component Fermi gas
with large scattering length has a tail
proportional to $1/k^4$ at large $k$. We show that the magnitude of this tail
is equal to the adiabatic derivative of the energy with respect to the reciprocal
of the scattering length, multiplied by a simple constant. This result holds at any temperature
(as long as the effective interaction radius is negligible)
and any large scattering length; it also applies to few-body cases. We then show some more
connections between the $1/k^4$ tail and various physical quantities, in particular the rate
of change of energy in a \emph{dynamic} sweep of the inverse scattering length.
\end{abstract}
\maketitle

\section{\label{sec:intro}The Theorem}
Consider an arbitrary number of fermions with mass $m$ in two spin states $\up$ and $\down$, having
an s-wave contact interaction with large scattering length $a$ ($\lvert a\rvert\gg
r_0$ \cite{_r_0}) between oppositite spin states, and confined by any external
potential $V_\text{ext}(\vect r)$. To discuss their momentum distribution, it is
convenient to use a box with volume $\Omega$ and impose a periodic boundary
condition. For a large uniform gas, $\Omega$ is its actual volume; but for a gas confined
in a trap, it is better to use (conceptually) a very large
box which contains the region of the gas. Now consider any stationary state, that is, any energy level
or any incoherent mixture of various energy levels (such
as an equilibrium state at any temperature \cite{_anyT}).
The momentum distribution $n_{\vect k\sigma}$ ($\leq1$ because of Pauli exclusion),
that is, the average number of fermions with wave vector $\vect k$ and spin $\sigma$,
has in general a well-known tail of the form $n_{\vect k\sigma}\approx C/k^4$
at large $\vect k$, where
\begin{equation*}
C\equiv\lim_{\vect k\rightarrow\infty}k^4n_{\vect k\sigma}
\end{equation*}
is independent of $\sigma$ \cite{TanEnergyTheorem}. In this equation, the limit
is physically restricted to $k\ll 1/r_0$, where $r_0$ is the effective interaction radius
\cite{_r_0}.

This tail determines how many large-momentum fermions (on average) we can find, if we measure
the momentum distribution of the gas. This measurement is practically
possible and has already been carried out to some extent \cite{Greiner2005PRL, Regal2005}
in some recent experiments
on the BCS (Bardeen-Cooper-Schrieffer superfluid)
to BEC (Bose-Einstein condensate) crossover behavior of such a Fermi gas \cite{crossover}.
The number of fermions with momenta larger than $\hbar K$ is, on average,
\begin{equation*}
N_{(k>K)}=\frac{\Omega C}{\pi^2K}
\end{equation*}
plus higher order corrections which are negligible at large $K$.
Here $K\gg 1/l_{\up\down}$, $K\gg1/\lambda_{dB}$, but $K\ll1/r_0$;
$l_{\up\down}$ is the characteristic distance between two fermions in the opposite spin states
\cite{_l_updown},
and $\lambda_{dB}\sim\hbar/\sqrt{mk_{B}T_\text{temp}}$ is the thermal de Broglie wave length.
$\hbar$ is Planck's constant divided by $2\pi$, $k_B^{}$ is Boltzmann's constant, and $T_\text{temp}$
is temperature \cite{_anyT}.

If we tune the reciprocal of the scattering length \cite{_tune} adiabatically (very slowly), 
but keep the confinement potential [such as a trap or an optical lattice
(for ultracold fermionic atoms)]
 fixed, the total energy $E$ (the sum of the kinetic energy,
the interaction energy, and the external trapping energy) of the 
gas will change accordingly, and we have a quantity $\dif E/\dif(1/a)$, which we call the adiabatic
derivative of energy (with respect to the inverse scattering length $1/a$).

The above two quantities, the magnitude of the $1/k^4$ tail, and the adiabatic derivative of
energy, can be independently measured in experiment.

We show in this paper that these two quantities have a \emph{simple},
\emph{exact} and \emph{universal} relation:
\begin{equation}\label{eq:adiabatic}
\frac{\hbar^2\Omega C}{4\pi m}=\frac{\dif E}{\dif (-1/a)},
\end{equation}
and we shall call it \emph{adiabatic sweep theorem}.

This theorem holds whenever the effective interaction radius $r_0$ \cite{_r_0}
 is negligible, compared to the other
relevant length scales of the problem: the scattering length, the average interparticle spacing,
the thermal de Broglie wave length, and the characteristic length scale over which the external potential
is inhomogeneous.
In this limit ($r_0\rightarrow 0$), the interaction between
two fermions in opposite spin states is s-wave contact interaction, characterized by the scattering
length only, and there is no interaction between fermions in the same spin state because
of Pauli blocking.

The theorem applies to any stationary state, such as
the ground state, any excited state, and any finite temperature state \cite{_anyT} in thermal
equilibrium.
Generally speaking, if $\ket{\phi_i}$ are a set of energy levels of the system, then
any mixed state described by a density operator $\rho=\sum_i\rho_i\ket{\phi_i}\brac{\phi_i}$
satisfies this theorem. The numbers of fermions in the two spin states $N_\up$ and $N_\down$,
the external potential (as long as it is fixed), and the scattering length are all arbitrary,
provided that the conditions specified in the previous paragraph are satisfied.

\section{\label{sec:proof}The proof}

Although the theorem can be proved with conventional means as well, we shall use the formalism
developed in \cite{TanEnergyTheorem}. This formalism, as we shall see below, allows us
to prove this universal theorem in a universal (and simple) way.

The basis of the formalism developed in \cite{TanEnergyTheorem}
 is a pair of linearly independent generalized
functions $\Lambda(\vect k)$ and $L(\vect k)$, satisfying
\begin{subequations}
\begin{eqnarray}
\Lambda(\vect k)&=1~~~(\lvert\vect k\rvert<\infty),~~~
\int\frac{\dif^3k}{(2\pi)^3}\frac{\Lambda(\vect k)}{k^2}&=0,\\
L(\vect k)&=0~~~(\lvert\vect k\rvert<\infty),~~~
\int\frac{\dif^3k}{(2\pi)^3}\frac{L(\vect k)}{k^2}&=1,\\
\Lambda(-\vect k)&=\Lambda(\vect k),~~~~~~~~~~~~~~~~~~~~~L(-\vect k)&=L(\vect k).
\end{eqnarray}
\end{subequations}
There is no contradiction at all, because the integrals of these functions over the whole
momentum space are \emph{not} solely determined by their values at \emph{finite} $\vect k$'s. Like
the delta function $\delta(x)$, which has a singularity at $x=0$, the above functions
have appropriate singularities at infinite momenta such that the above integrals hold.

This formalism is equivalent to the one based on the Fermi pseudopotential
$4\pi\hbar^2a\delta(\vect r)(\partial/\partial r)r/m$ \cite{HuangYang, LeeHuangYang, _FermiPseudopotential},
but it has a greater
simplicity and flexibility, making concrete calculations more straightforward, and
allowing for some useful formal derivations. The present author
has used this new formalism to find another exact theorem \cite{TanEnergyTheorem}, and solve the
two-body and three-body problems \cite{TanEnergyTheorem}. The dimer-fermion scattering length
calculated by the author \cite{TanEnergyTheorem}
is completely consistent with independent calculations by others \cite{Skorniakov, Petrov2005PRA},
but has a much higher accuracy \cite{TanEnergyTheorem}. In addition, the author used it to
calculate the dimer-dimer scattering length (unpublished), and got $a_{dd}=0.6a$, consistent with
\cite{Petrov2004PRL}.
The author also used this formalism to study 
the three-boson problem with large scattering length, the beyond-mean-field ground state
of the dilute Bose gas, and the ground state of the dilute two-component
Fermi gas with short range interaction. In \emph{all} the instances, the results
obtained within this seemingly paradoxical formalism are perfectly consistent with the celebrated
results in the literature. Finally, the formalism and the theorem derived in \cite{TanEnergyTheorem}
have been easily generalized to two dimensions \cite{Tan2D}.

In the language of these generalized functions, the system's hamiltonian is
\begin{equation}
H(a)=T+aU+V,
\end{equation}
where the three operators
\begin{align*}
T&=\sum_{\vect k\sigma}(\hbar^2 k^2/2m)c_{\vect k\sigma}^\dagger c_{\vect k\sigma},\\
U&=\frac{4\pi\hbar^2}{m\Omega}\sum_{\vect q\vect k\vect k'}
c_{\vect q/2+\vect k\up}^\dagger
c_{\vect q/2-\vect k\down}^\dagger c_{\vect q/2-\vect k'\down}^{}
c_{\vect q/2+\vect k'\up}^{}\Lambda(\vect k'),\\
V&=\sum_\sigma\int\dif^3r V_\text{ext}(\vect r)
\psi_{\sigma}^\dagger(\vect r)\psi_{\sigma}^{}(\vect r)
\end{align*}
are all independent of $a$. Here $c_{\vect k\sigma}^{}$ and $\psi_\sigma^{}(\vect r)$
are the conventional fermionic annihilation operators in momentum and coordinate representations,
respectively.

Any energy level $\ket{\phi}$ associated with the hamiltonian $H(a)$, such that
$H(a)\ket{\phi}=E\ket{\phi}$,
automatically satisfies a short-range boundary condition \cite{TanEnergyTheorem}
\begin{equation}\label{eq:boundarycondition}
\sum_{\vect k}\eta(\vect k)c_{\vect q/2-\vect k\down}^{}c_{\vect q/2+\vect k\up}^{}\ket{\phi}=0
~~~~(\text{any $\vect q$}),
\end{equation}
where $\eta(\vect k)\equiv\Lambda(\vect k)+ L(\vect k)/4\pi a$.
This is just a rephrasing \cite{TanEnergyTheorem, Tan2D}
of the familiar statement that the wave function is of the form
$\text{coeff.}(1/r-1/a)+O(r)$ when the distance $r$ between two fermions in opposite spin states
approaches zero.

Further, it is shown that any state $\ket{\phi}$ which satisfies
Eq.~\eqref{eq:boundarycondition} must satisfy an energy theorem \cite{TanEnergyTheorem}:
\begin{equation}
\brac{\phi}H(a)\ket{\phi}=\brac{\phi}(\eta T+V)\ket{\phi},
\end{equation}
where $\eta T$ is the shorthand for
$\sum_{\vect k\sigma}\eta(\vect k) (\hbar^2k^2/2m)c_{\vect k\sigma}^\dagger
c_{\vect k\sigma}^{}$.

If we thoroughly
examine the logic in the long proof of this energy theorem in \cite{TanEnergyTheorem},
we find that the symbol $\eta$ is only associated with the boundary condition satisfied
by $\brac{\phi}$, but has nothing to do with $\ket{\phi}$; that is, we can consider
any energy eigenstate $\ket{\phi}$ of $H(a)$ [such that $H(a)\ket{\phi}=E\ket{\phi}$]
and any energy eigenstate $\ket{\phi'}$ of \emph{another} hamiltonian $H(a')$ [such that
$H(a')\ket{\phi'}=E'\ket{\phi'}$], where $H(a')$ differs from $H(a)$ only in the scattering
length, and prove that
\begin{equation}\label{eq:EnergyTheoremExtended}
\brac{\phi}H(a')\ket{\phi'}=\brac{\phi}(\eta T+V)\ket{\phi'}
\end{equation}
using the same logic as in the proof of the energy theorem.

We may also follow another approach in \cite{TanEnergyTheorem}, and derive this equality much more easily:
$\brac{\phi}(T+a'U+V)\ket{\phi'}=
\brac{\phi}(\eta T+a'\eta U+V)\ket{\phi'}
=\brac{\phi}(\eta T+V)\ket{\phi'}$,
where $\eta U$ is the shorthand for
$\frac{4\pi\hbar^2}{m\Omega}\sum_{\vect q\vect k\vect k'}\eta(\vect k)
c_{\vect q/2+\vect k\up}^\dagger
c_{\vect q/2-\vect k\down}^\dagger c_{\vect q/2-\vect k'\down}^{}
c_{\vect q/2+\vect k'\up}^{}\Lambda(\vect k')$, and $\brac{\phi}\eta U=0$
because of the boundary condition $\brac{\phi}\sum_{\vect k}\eta(\vect k)c_{\vect q/2+\vect k\up}^\dagger
c_{\vect q/2-\vect k\down}^\dagger=0$.

Now Eq.~\eqref{eq:EnergyTheoremExtended} can be further simplified as follows, because
$\ket{\phi'}$ is an eigenstate of $H(a')$:
\begin{equation}\label{eq:phi_phi'}
E'\bracket{\phi}{\phi'}=\brac{\phi}(\eta T+V)\ket{\phi'}.
\end{equation}

Similarly, we have proved that
\begin{equation}\label{eq:phi'_phi}
E\bracket{\phi'}{\phi}=\brac{\phi'}(\eta' T+V)\ket{\phi},
\end{equation}
where $\eta'(\vect k)\equiv\Lambda(\vect k)+L(\vect k)/4\pi a'$.

Taking the complex conjugate of Eq.~\eqref{eq:phi'_phi}, noting that
both $\eta' T$ and $V$ are hermitian, and subtracting the result from
Eq.~\eqref{eq:phi_phi'}, we get $(E'-E)\bracket{\phi}{\phi'}=\brac{\phi}(\eta-\eta')T\ket{\phi'}$,
and both the operator $V$ and the function $\Lambda(\vect k)$ are canceled. What remains is
\begin{multline}\label{eq:phi_phi'dE}
(E'-E)\bracket{\phi}{\phi'}\\=(1/4\pi a-1/4\pi a')
\brac{\phi}
\sum_{\vect k\sigma}L(\vect k)\frac{\hbar^2k^2}{2m}c_{\vect k\sigma}^\dagger c_{\vect k\sigma}^{}\ket{\phi'}.
\end{multline}

Now we apply Eq.~\eqref{eq:phi_phi'dE} to a special situation,
in which we do an adiabatic sweep of the inverse scattering length from $1/a$ to $1/a'$,
and the state $\ket{\phi}$ is continuously transformed to $\ket{\phi'}$ (that is, $\ket{\phi'}$ is adiabatically
connected to $\ket{\phi}$). In the limit $1/a'\rightarrow1/a$, such that
$\ket{\phi'}\rightarrow\ket{\phi}$, we obtain a differential relation
\begin{equation}\label{eq:differential}
-\frac{\dif E}{\dif(1/a)}=\frac{\hbar^2}{8\pi m}\sum_{\vect k\sigma}L(\vect k)k^2
\brac{\phi}c_{\vect k\sigma}^\dagger c_{\vect k\sigma}^{}\ket{\phi}/\bracket{\phi}{\phi}.
\end{equation}

By the basic property of $L(\vect k)$ \cite{TanEnergyTheorem},
$\sum_{\vect k}L(\vect k)f(\vect k)=\Omega\int[\dif^3k/(2\pi)^3]L(\vect k)f(\vect k)
=\Omega\lim_{\vect k\rightarrow\infty}k^2f(\vect k)$ for any conventional function
$f(\vect k)$. So the summation in Eq.~\eqref{eq:differential} is equal to
$\Omega\sum_\sigma\lim_{\vect k\rightarrow\infty}k^4\langle c_{\vect k\sigma}^\dagger
c_{\vect k\sigma}^{}\rangle=2\Omega C$. Therefore,
\begin{equation*}
-\frac{\dif E}{\dif(1/a)}=\frac{\hbar^2\Omega C}{4\pi m},
\end{equation*}
and we have proved the adiabatic sweep theorem in the case of any pure energy eigenstate.

In any incoherent mixed state of the form $\rho=\sum_i\rho_i\ket{\phi_i}\brac{\phi_i}$, where
$\ket{\phi_i}$ are energy eigenstates, both $E$ and $C$ are statistical averages of the $E$
and $C$ values in the states $\ket{\phi_i}$ with weights $\rho_i$, so the adiabatic sweep theorem
obviously still holds.

\section{Two-body case: a conventional proof}
It is cumbersome to prove the above theorem with conventional means. Here we just demonstrate the two-body
case, and then sketch the extension to the $N$-body cases. This shall help to convince
the readers that the adiabatic sweep theorem is indeed correct, even if they are not
used to the succinct (and ``unconventional'') proof in the last section.

Now consider two fermions in opposite spin states with an external potential $V_\text{ext}(\vect r)$.
We consider an energy eigenstate described by a wave function $\phi(\vect r_1, \vect r_2)$, with scattering
length $a$ and energy $E$; we also consider another energy eigenstate described by a wave function
$\phi'(\vect r_1, \vect r_2)$, with scattering length $a'$ and energy $E'$. Here  $\vect r_1$ is the position
of a spin up fermion, and $\vect r_2$, spin down.
\begin{equation}
\left[-\frac{\hbar^2\nabla_1^2}{2m}-\frac{\hbar^2\nabla_2^2}{2m}+V_\text{ext}(\vect r_1)
+V_\text{ext}(\vect r_2)\right]\phi=E\phi
\end{equation}
whenever $\vect r_1\neq\vect r_2$. There is also a short-range boundary condition:
\begin{equation}\label{eq:short-range2body}
\phi(\vect r_1, \vect r_2)=A(\vect r_0)(1/r-1/a)+O(r)~~~~(r\rightarrow 0),
\end{equation}
where $\vect r_0\equiv(\vect r_1+\vect r_2)/2$ and $\vect r\equiv\vect r_1-\vect r_2$.
Similar equations hold for $\phi'$. It then follows that
\begin{equation}
-\frac{\hbar^2}{2m}\sum_{i=1}^{2}\nabla_i\cdot(\phi'^*\nabla_i\phi-\phi\nabla_i\phi'^*)=(E-E')\phi'^*\phi.
\end{equation}
Making the substitution $\nabla_1=\nabla_0/2+\nabla_\vect{r}$ and $\nabla_2=
\nabla_0/2-\nabla_\vect{r}$ (where $\nabla_0$ and $\nabla_\vect{r}$ are partial differential
operators associated with $\vect r_0$ and $\vect{r}$, respectively), we get
\begin{multline}
-\frac{\hbar^2}{4m}\Bigl[\nabla_0\cdot(\phi'^*\nabla_0\phi-\phi\nabla_0\phi'^*)\\
+4\nabla_\vect{r}\cdot(\phi'^*\nabla_\vect r\phi-\phi\nabla_\vect{r}\phi'^*)\Bigr]
=(E-E')\phi'^*\phi.
\end{multline}
We then integrate this equation over $\mathcal{D}_\epsilon$ ($\epsilon>0$), the set of all configurations
$(\vect r_1, \vect r_2)$ in which $r>\epsilon$. The left side is converted to
surface-flux according to Stokes theorem. Since
the wave functions either decay to zero at infinity or satisfy a periodic boundary condition in a box,
only the flux on the hypersurface $r=\epsilon$ remains. Using the short-range boundary conditions,
such flux is found to be $(4\pi\hbar^2/m)(1/a'-1/a)\int\dif^3r_0A'^*(\vect r_0)A(\vect r_0)$ plus
correction terms which vanish in the limit $\epsilon\rightarrow0^+$. Taking the limit $\epsilon\rightarrow
0^+$, we get
\begin{multline}\label{eq:twobodyonly}
(E-E')\int\dif^3r_1\dif^3r_2\phi'^*\phi\\=(4\pi\hbar^2/m)
(1/a'-1/a)\int\dif^3r_0A'^*(\vect r_0)A(\vect r_0).
\end{multline}
Now we specialize to the case in which $\phi'$ is adiabatically connected to $\phi$, and take the limit
$1/a'\rightarrow 1/a$; we further assume that $\phi$ is normalized. The above equation is then
simplified as
\begin{equation}\label{eq:difE}
-\frac{m}{4\pi\hbar^2}\frac{\dif E}{\dif(1/a)}=\int\dif^3r_0\lvert A(\vect r_0)\rvert^2.
\end{equation}

Now we turn to the evaluation of $\Omega C$ in the quantum state $\phi$.
We need to transform the wave function to momentum space:
\begin{equation}
\widetilde{\phi}_{\vect k_1\vect k_2}=\int\frac{\dif^3r_1\dif^3r_2}{\Omega}\phi(\vect r_1\vect r_2)
\exp(-\I\vect k_1\cdot\vect r_1-\I\vect k_2\cdot\vect r_2),
\end{equation}
and the average number of spin-up fermions with wave vector $\vect k$ is
\begin{equation}\label{eq:nk}
n_{\vect{k}\up}=\sum_{\vect k'}\bigl\lvert\widetilde{\phi}_{\vect k\vect k'}\bigr\rvert^2.
\end{equation}
At large $\vect k$, $\widetilde{\phi}_{\vect k\vect k'}$ is dominated by the contribution
from the most singular term in $\phi(\vect r_1\vect r_2)$
(it is a general property of Fourier transformation), namely $A(\vect r_0)/r$. So
\begin{equation}
\widetilde{\phi}_{\vect k\vect k'}=\Omega^{-1}\widetilde{A}(\vect k+\vect k')16\pi/(\vect k-\vect k')^2
\end{equation}
plus higher order terms, at large $\vect k$. Here $\widetilde{A}(\vect q)\equiv\int\dif^3r_0A(\vect r_0)
\exp(-i\vect q\cdot \vect
r_0)$. Note that $A(\vect r_0)$ describes the center of mass motion of the two fermions
(when their distance is zero) and should be
 everywhere finite and continuous; consequently, $\widetilde{A}(\vect q)$ should decay
sufficiently fast at large $\vect q$. It then follows that the summation in Eq.~\eqref{eq:nk} should
be dominated by terms in which $\lvert\vect k+\vect k'\rvert\ll k$, at large $\vect k$. So
\begin{equation}
n_{\vect k\up}=\sum_{\vect q}\bigl[\Omega^{-1}\widetilde{A}(\vect q)4\pi/k^2\bigr]^2
\end{equation}
plus higher order corrections, at large $\vect k$. So
\begin{equation}\label{eq:C}
C=\lim_{\vect k\rightarrow\infty}k^4 n_{\vect k\up}=\Omega^{-1}16\pi^2\int\dif^3r_0\lvert A(\vect r_0)\rvert
^2,
\end{equation}
and the same formula results if we study the spin-down fermion.
Comparing this with Eq.~\eqref{eq:difE}, we find
\begin{equation*}
\frac{\hbar^2\Omega C}{4\pi m}=-\frac{\dif E}{\dif(1/a)}.
\end{equation*}

If the two fermions are in the same spin state, there is no interaction between them, and $E$ is independent
of $a$; meanwhile, $C=0$.

The conventional proof of Eq.~\eqref{eq:adiabatic} for more fermions is similar to that of the two-body
case. The major additional considerations are:
\begin{itemize}
\item In the intermediate steps, the set $\mathcal{D}_\epsilon$ (see above) also
excludes the configurations in which three or more fermions cluster in a tiny region of size $\epsilon$.
When we apply Stokes
theorem, we also need to consider surface contributions from these excluded configurations. But Pauli
blocking suppresses the probability amplitude for three or more fermions to cluster in such a tiny region.
It then follows that surface contributions from these three-body and more-body clusters
vanish in the limit $\epsilon\rightarrow 0$, leaving us with contributions from the two-body clusters only.
An equation similar to Eq.~\eqref{eq:twobodyonly} thus results. 
\item Three-body or more-body clusters can only contribute a tail of the form $O(1/k^\alpha)$
($\alpha>5$) to the momentum distribution at large $k$, because of Pauli blocking \cite{_tail}. 
The $1/k^4$ tail is solely due to the two-body (opposite spin) clusters, and an equation
similar to Eq.~\eqref{eq:C} still holds.
\end{itemize}

\section{\label{sec:implications}The Implications}
\subsection{Physical meaning}
The meaning of Eq.~\eqref{eq:adiabatic} is intuitively clear.  If we tune the
inverse scattering length slightly and slowly, quantum mechanical 
first-order perturbation determines the amount by which the energy level shifts.
In the limit of zero effective interaction radius,
two fermions (in opposite spin states) do not interact unless they appear at the same position,
so the energy shift should be proportional to the probability that this occurs.
Meanwhile, $\Omega C$ characterizes this probability \cite{TanEnergyTheorem};
recall that when two fermions appear at the same place, their momenta are large and nearly opposite.
The definitive form of Eq.~\eqref{eq:adiabatic} is of course not obvious.

\subsection{Large momentum part in the BEC-BCS crossover}
In this subsection we are restricted to a special (but interesting)
 case: the uniform Fermi gas with equal
populations of the two spin states \emph{at zero temperature}. The scattering length $a$ is large
($\lvert a\rvert\gg r_0$) and arbitrary.

There is only one independent dimensionless parameter, $x\equiv-1/k_F^{}a$,
where
\begin{equation*}
k_F^{}\equiv(3\pi^2n)^{1/3}
\end{equation*}
is the Fermi wave vector,
and $n$ is the number density of fermions.
When $x$ is negative and large, the ground state \cite{_ground_state}
is a BEC of fermionic dimers \cite{crossover};
when $x$ is positive and large, the gas becomes a BCS superfluid \cite{crossover};
in between, the system is in a strongly correlated many-body state,
but there is no quantum phase transition \cite{crossover}. For all finite $x$ values (including $x=0$),
the ground state is believed to be a superfluid \cite{crossover}.

According to Eq.~\eqref{eq:adiabatic}, we just need to take the derivative of the ground state energy
with respect to $-1/a$, to get the magnitude of the large momentum part, the contact intensity
\cite{TanEnergyTheorem} $C$.

\textbf{\emph{BEC limit--}} the ground state energy is \cite{LeeYang, LeeHuangYang, Wu1959PR,
Braaten2002PRL, Petrov2004PRL, Stringari2004Eu, TanLevin}
\begin{multline}
E/\Omega=n_dE_d+2\pi\hbar^2 n_d^2 a_{dd}/m_d\\
\times\bigl(1+4.8144\sqrt{g}+19.6539g\ln g+c_3^{} g+\cdots\bigr),
\end{multline}
where $n_d=n/2$ is the dimers' number density, $E_d=-\hbar^2/ma^2$ is the energy of an isolated
dimer at rest, $m_d=2m$ is the dimer's mass, $g=n_da_{dd}^3$ is the Bose gas parameter.
$a_{dd}=c_\text{P}^{}a$ \cite{Petrov2004PRL}
 is the dimer-dimer scattering length, and $c_\text{P}^{}\approx0.6$ \cite{Petrov2004PRL}.
The constant $c_3$ is still unknown, since it depends on the difficult three-dimer scattering
problem \cite{Braaten2002PRL, TanLevin}. The contact intensity thus has an expansion
\begin{multline}\label{eq:C_BEC}
C=4\pi n/a+c_\text{P}^{}\pi^2n^2a^2\bigl[1+12.0360\sqrt{g}\\
+78.6156g\ln g+(4c_3^{}+58.9617)g+\cdots\bigr].
\end{multline}

\textbf{\emph{Unitary regime--}}
$-1/k_F^{}a$ is small or zero (the latter case is called unitarity limit \cite{Baker, Ohara, Ho}).
 There is an expansion \cite{Baker, Ohara, Ho, Chang2004PRA,
Astrakharchik2004PRL, Bulgac2005PRL}
\begin{equation}\label{eq:E_unitary}
\frac{E}{\Omega}
=\frac{3}{5}\frac{\hbar^2k_F^2n}{2m}\biggl[\xi-\frac{\zeta}{k_F^{}a}-\frac{5v}{3(k_F^{}a)^2}+\cdots\biggr],
\end{equation}
where the current theoretical estimate of the constants are
$\xi\approx0.44$ \cite{Chang2004PRA, Astrakharchik2004PRL}, $\zeta\approx1$ 
\cite{Chang2004PRA, Astrakharchik2004PRL, Bulgac2005PRL} and $v\approx1$ \cite{Chang2004PRA, Bulgac2005PRL}.
Consequently,
\begin{equation}\label{eq:C_unitary}
C=k_F^4\biggl(2\zeta/5\pi+\frac{4v/3\pi}{k_F^{}a}+\cdots\biggr).
\end{equation}
In particular, in the unitarity limit,
\begin{equation}
C=jk_F^4  ~~~~(\text{unitarity limit}),
\end{equation}
where
\begin{equation}j=2\zeta/5\pi\end{equation}
is a universal constant. \emph{If} $\zeta\approx1$
\cite{Chang2004PRA, Astrakharchik2004PRL, Bulgac2005PRL}, $j\approx0.13$;
this value of $j$ is consistent with a recent Monte Carlo calculation of the full momentum distribution
\cite{Astrakharchik2005}. Of course, if $\zeta$ is appreciably different from 1, $j$ must be appreciably different
from $0.13$.

\textbf{\emph{BCS limit--}}
the ground state energy is \cite{Baker}
\begin{multline}\label{eq:E_BCS}
\frac{E}{\Omega}=\frac{\hbar^2k_F^2n}{m}\biggl[\frac{3}{10}+\frac{1}{3\pi}k_F^{}a+0.055661(k_F^{}a)^2\\
+0.00914(k_F^{}a)^3-0.018604(k_F^{}a)^4+\cdots\biggr],
\end{multline}
where $2(11-2\ln2)/35\pi^2\approx0.055661$ is exact \cite{LeeYang},
but the last two coefficients still have uncertainties \cite{Baker}; the BCS superfluid energy saving
is exponentially small [$\sim\exp(-\pi/2k_F^{}\lvert a\rvert)$] (see for example \cite{deGennes})
and does not show up in this asymptotic power series of $k_F^{}a$ \cite{Astrakharchik2004PRL}. So
\begin{multline}\label{eq:C_BCS}
C=4\pi^2n^2a^2\Bigl[1+1.049190k_F^{}a\\+0.2584(k_F^{}a)^2-0.70135(k_F^{}a)^3+\cdots\Bigr],
\end{multline}
where $12(11-2\ln2)/35\pi\approx1.049190$.

The full momentum distribution in the BCS limit has actually been calculated by Belyakov \cite{Belyakov}
and later corrected by Sartor \textit{et al} \cite{Sartor}, to the lowest nontrivial order 
[$O(a^2)$] in the scattering length only,
although these authors studied the problem in the context of a dilute hard sphere Fermi gas
(with positive scattering length), and they did not foresee the applicability of their result
to a dilute ($1/k_F^{}\gg\lvert a\rvert$) Fermi gas with a large ($\lvert a\rvert\gg r_0$)
and \emph{negative} scattering length \cite{_BCS_blurring}. The universal momentum distribution function
[to $O(a^2)$] they \cite{Belyakov, Sartor}
 derived has a tail consistent with the leading order term in Eq.~\eqref{eq:C_BCS}.

Note that $C\propto n^2a^2$ in the BCS limit, in contrast with the predictions based on the BCS
equations, which lead to an exponentially small value [$\sim\exp(-\pi/k_F^{}\lvert a\rvert)$]
of $C$.
\emph{BCS equations do not predict the correct behavior of $C$ in the BCS limit.}
This observation is not incompatible with a recent experiment \cite{Regal2005},
in which the measured large momentum tail of a trapped Fermi gas in the BCS regime is found \cite{Regal2005} to
be stronger than the theoretical prediction based on the BCS theory.

\begin{figure}
\includegraphics{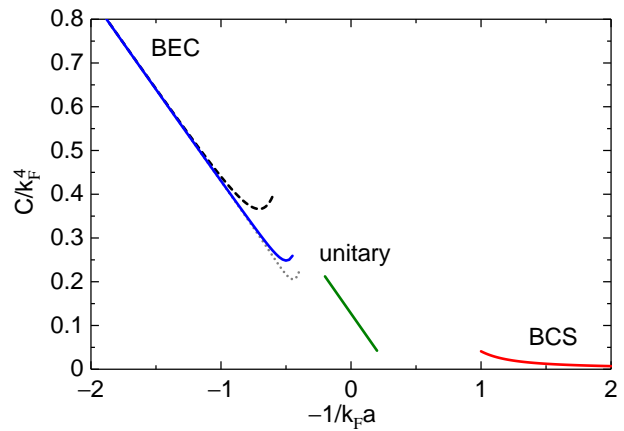}
\caption{\label{fig:ContactIntensity}(color online) Intensity of the large
 momentum part $C\equiv\lim_{\vect k\rightarrow\infty}
k^4n_{\vect k\sigma}^{}$ in the BEC-BCS crossover at zero temperature.
BEC regime: Eq.~\eqref{eq:C_BEC},
in which the dashed line is for $c_3^{}=140$ \cite{Braaten2002PRL, TanLevin},
solid line for $c_3^{}=40$ \cite{TanLevin},
and gray dotted line, $c_3^{}=30$ \cite{TanLevin} (the upward bending
of each of these curves toward the unitary regime marks the breakdown of
the low-density expansion).
Unitary regime: Eq.~\eqref{eq:C_unitary} combined with the numbers $\zeta\approx1$ and $v\approx1$ 
(note that the actual slope of the curve in
the unitarity limit is probably less steep; see text).
BCS regime: Eq.~\eqref{eq:C_BCS}, which should be less reliable when $-1/k_F^{}a\approx1$.
}
\end{figure}

These various expansions of $C$ are plotted in Fig.~\ref{fig:ContactIntensity}.
It is almost certain that $C$ monotonically decreases from the BEC limit to the BCS limit, at a fixed particle
density. The parameter $v$ determines the slope of the $C$ curve in the unitarity limit ($-1/k_F^{}a=
0$), and it seems that the value $v\approx1$ (responsible for the slope of the short line
in the unitary regime) is probably too large.

What we have learned here, which is based on Eq.~\eqref{eq:adiabatic}, is remarkable:
\emph{simply measuring the coefficient of the $1/k^4$ tail at various scattering lengths,
one could pin down the whole equation of state of this novel Fermi gas} \cite{_BEC_caveat}.

Even the famous constant [see Eq.~\eqref{eq:E_unitary}]
\begin{equation*}
\xi\equiv1+\beta,
\end{equation*}
characterizing the ground state energy of the unitary Fermi gas \cite{Baker, Ohara},
can be alternatively determined from measurements
of $C$ \emph{away from the unitarity limit}. Suppose that 
\begin{equation*}
C=k_F^4c(-1/k_F^{}a),
\end{equation*}
and we can show from Eq.~\eqref{eq:adiabatic} that
\begin{equation}
\beta=-\frac{5\pi}{2}\int_0^{+\infty} c(x)\dif x.
\end{equation}
$c(x)$ also must satisfy a constraint [according to Eq.~\eqref{eq:adiabatic} and
the behavior of the energy density in the BEC and BCS limits]:
\begin{equation}
\int_{-\infty}^0\bigl[c(x)+4x/3\pi\bigr]\dif x
+\int_0^{+\infty}c(x)\dif x=\frac{2}{5\pi}.
\end{equation}

We finally remark on the \emph{practical accessibility} of the $C/k^4$ tail in the momentum distribution.
In the BCS limit, the result in \cite{Belyakov, Sartor} indicates that
the \emph{difference} between $k^4n_{\vect k\sigma}$
and $C$ is already as small as a few percent, when $k/k_F^{}=5$, and it decreases
fast at larger $k$ values. The situation is similar in the unitary regime, as
indicated by a recent Monte-Carlo simulation \cite{Astrakharchik2005}.
In the BEC limit, we expect the difference between $k^4n_{\vect k\sigma}$ and the true
value of $C$ to be negligible when $k$ is more than a few times $1/a$ ($a$ characterizes
the size of each fermionic dimer).

\subsection{Large momentum part of the trapped gas}
For the gas confined in a trap, the parameter $C$ in Sec.~\ref{sec:intro} is only the \emph{average}
contact intensity over the arbitrary volume $\Omega$. It is the integrated contact intensity,
$\Omega C$, that is uniquely defined.

We can, however, uniquely define the \emph{local contact intensity}
\begin{equation}
C(\vect r)\equiv-4\pi\lim_{s\rightarrow0^+}s\nabla_{\vect s}^2\langle
\psi_\sigma^\dagger(\vect r)\psi_\sigma^{}(\vect r+\vect s)\rangle,
\end{equation}
which is also independent of $\sigma$. The limit is restricted to $s\gg r_0$. It is
straightforward to show that
\begin{equation}
\Omega C=\int\dif^3rC(\vect r).
\end{equation}

An advantage of $C(\vect r)$ is that for a trapped cloud of many fermions, we can
use the local density approximation to predict the value of $\Omega C$ from the
equation of state of the \emph{uniform} gas. Let us illustrate this with the zero temperature
case. First, the local density of the gas
is estimated from the equation
$\mu_\text{local}[n(\vect r)]+V_\text{ext}(\vect r)=$const, where $\mu_\text{local}(n)$
is the chemical potential of the uniform gas with density $n$.
Then $C(\vect r)$ is approximately calculated from this local value of $n$
using the formula of $C$ for the uniform gas. Finally we do the spatial integral
and predict (approximately) the intensity of the large momentum part of the trapped gas.
Note that in this analysis, the formula for $C$ of the uniform gas is also solely
determined by the equation of state.

By comparing our prediction with the measured large momentum part, we could find out
whether a particular equation of state we have used in this kind of analysis is correct
and, if not, we could adjust it. In this way, the experimental data on the large momentum
part may also help us to pin down the equation of state of the uniform gas.

We now turn to another question: how to \emph{directly test} the adiabatic sweep theorem in a trap
at any temperature \cite{_anyT}?
At least one approach is feasible in principle
(for $1/a\leq0$ only) and involves almost no theoretical approximations.
1) We first prepare a sample of equilibrated Fermi gas in a particular trap
 and then take \textit{in situ} image of the gas to determine its external
trapping energy. 2) We produce the trapped sample again, following exactly the same recipe as before,
but then suddenly switch off the trap (without changing the scattering length) to allow the gas
to expand; after enough time, an image is taken, and the kinetic plus interaction energy
of the gas is deduced from the second moment of the spatial profile. The sum of the above two energies
is the total energy. 3) The sample is prepared the third time, and then the inverse scattering length
is adiabatically tuned (over a time scale longer than the relaxation time)
to a somewhat different value $1/a'$, and then the external trapping energy is measured in the same way
as step one. 4) The sample is prepared and the inverse scattering length adiabatically tuned to $1/a'$
again, and the kinetic plus interaction energy is measured like in step two. The sum of the energies
from steps three and four is the energy of the sample at that new scattering length,
which should be different from the sum of energies from steps one and two. This difference,
according to our theorem, must determine the intensity of the $1/k^4$ tail (if the new inverse
scattering length does not differ too much from the old one). So we can prepare
the sample for the fifth time, measure its momentum distribution, and compare the large momentum part
with the prediction based on our theorem.

The above five-step approach (if somewhat tedius) should be robust. A variant of this approach
(which may be more practical) is to measure the total energy at a series of $-1/a$ values
(all of the states are adiabatically connected), do a curve fit, and take the derivative 
of the curve with respect to $-1/a$. In this way, the adiabatic derivative of energy
can be measured more accurately. No local density approximation is involved in this approach.

\subsection{Thermodynamics at any temperature}
In this subsection we consider the \emph{uniform} gas at \emph{any} temperature \cite{_anyT}.
The two spin states are assumed to be equally populated.
Now we have two independent dimensionless variables,
$x\equiv-1/k_F^{}a$, and the average entropy per fermion $s\equiv S/k_B^{}N$.
Here $S$ is the entropy, and $N$ is the number of fermions.

The contact intensity is then determined by a universal dimensionless function $c(x,s)$:
\begin{equation}\label{eq:c_function}
C=k_F^4 c(x,s).
\end{equation}
\emph{This function contains all the information about 
the thermodynamic properties of this novel Fermi gas},
including the phase transition between the superfluid and normal states.
Here we use entropy per particle instead of the reduced temperature (temperature divided by the Fermi temperature),
because entropy is invariant in an adiabatic sweep and is thus a more convenient variable here.

The properties of this Fermi gas are known in each of these three limits:
$x\rightarrow-\infty$, $x\rightarrow+\infty$, and $s\rightarrow\infty$
\cite{Ho_highT}. The only difficult
region is at $x\sim O(1)$ and $s\sim O(1)$, abbreviated below as $D$.
If we start from the $x\rightarrow\pm\infty$ regions, and tune the inverse scattering length
adiabatically into $D$, we know the entropy of the final state from that of the well-known
initial state. Measuring $C$ at various $x$ values in between, we can integrate $C$ over $x$ to get the
energy of the final state. If we repeat the same measurements at various \emph{initial}
entropies, we can calculate the derivative of energy with respect to entropy at any scattering length,
and obtain the temperature of the gas in $D$.

This program is viable in principle, and it has the advantage of being model-independent
(it is needless to say that there are other experimental approaches to the thermodynamic quantities as well;
purely for the sake of space, we will not list or compare them in this paper),
since almost no theoretical approximations are involved \cite{_contact}.
In reality it is complicated by the fact that
samples of such Fermi gas studied to date are all in small traps
or in optical lattices, which have inhomogeneous density distribution. However using some ingenious data
analysis method, we might still be able to extract valuable information from experimental data
on the large momentum part.

We can also derive a useful relation about the \emph{pressure} of the uniform
Fermi gas with large scattering length and equal populations of the two spin states, in thermal
equilibrium:
\begin{equation}\label{eq:P1}
P=\frac{2}{3}\rho_E^{}+\frac{\hbar^2 C}{12\pi am},
\end{equation}
where $\rho_E^{}\equiv E/\Omega$ is the energy density. This relation holds at any temperature \cite{_anyT}.
\begin{proof}
On dimensional grounds, the energy density is $\rho_E^{}=\hbar^2k_F^5f(x,s)/m$, where $x=-1/k_F^{}a$ and $s$
is the average entropy per fermion. Using Eq.~\eqref{eq:adiabatic},
we can easily show that $c(x,s)=4\pi(\partial/\partial x)f(x,s)$, where $c(x,s)$ is defined
by Eq.~\eqref{eq:c_function}.

Suppose that the total number of fermions is $N$, and the volume
is $\Omega$. Now we compress the gas adiabatically (but keep the scattering length unchanged),
to reduce its volume to a slightly smaller value
$\Omega'\equiv(1-o)^3\Omega$. Consequently, the average interfermionic spacing $1/k_F$ is reduced by a factor
$1-o$, so $x'=(1-o)x$. The Fermi wave vector becomes
$k_F^{'}=k_F^{}/(1-o)$.
 So the energy
of the gas changes by $\Delta E=\Omega'\hbar^2k_F'^5f(x',s)/m-\Omega\hbar^2k_F^5f(x,s)/m$,
and the pressure is $P=-\lim_{o\rightarrow0}\Delta E/(\Omega'-\Omega)
=(\hbar^2 k_F^5/m)[2f/3-x(\partial/\partial x)f/3]=(\hbar^2 k_F^5/m)(2f/3-xc/12\pi)
=2\rho_E^{}/3+\hbar^2C/(12\pi am)$.
\end{proof}

If we further express the energy density in terms of the momentum distribution \cite{TanEnergyTheorem},
namely $\rho_E^{}=\hbar^2C/4\pi am+\sum_{\sigma}\int[\dif^3k/(2\pi)^3]\Lambda(\vect k)(\hbar^2k^2/2m)
n_{\vect k\sigma}$, we see that the pressure (in the equilibrium state)
is a simple functional of the momentum distribution
\begin{align}\label{eq:P2}
P&=\frac{\hbar^2C}{4\pi am}+\frac{2}{3}\lim_{K\rightarrow\infty}\sum_\sigma
\int_{\lvert\vect k\rvert<K}\frac{\dif^3k}{(2\pi)^2}\frac{\hbar^2k^2}{2m}
\Bigl(n_{\vect k\sigma}-\frac{C}{k^4}\Bigr)\nonumber\\
&=\frac{\hbar^2C}{4\pi am}+\frac{2}{3}\sum_\sigma\int\frac{\dif^3k}{(2\pi)^2}\Lambda(\vect k)
\frac{\hbar^2k^2}{2m}n_{\vect k\sigma}\nonumber\\
&=\sum_\sigma\int\frac{\dif^3k}{(2\pi)^3}\biggl[\frac{2}{3}\Lambda(\vect k)+\frac{L(\vect k)}{4\pi a}\biggr]
\frac{\hbar^2k^2}{2m}n_{\vect k\sigma},
\end{align}
where $K\ll1/r_0$ physically. We shall call this result \emph{pressure theorem}
(of the two-component Fermi gas with large scattering length).

\subsection{Dynamic sweep}
What if we tune the inverse scattering length at a fast rate, so that the
system's evolution is \emph{not} adiabatic?
Do we still have some insight into the rate of change
of the system's energy (or, more precisely, expectation value of energy, since the system is
not in any energy eigenstate if it is changing with time) \cite{_energy_change}?

If the characteristic time scale $t_\text{sweep}$ of the sweep is comparable
to or shorter than $mr_0^2/\hbar$ \cite{_mr02_example},
we have to either modify or completely give up the s-wave contact
interaction model; this topic will \emph{not} be discussed here.

If $t_\text{sweep}\gg mr_0^2/\hbar$ \cite{_mr02_example},
one can reasonably expect that the sweep can still be regarded
as adiabatic from the perspective of the short-range boundary condition, \textit{ie}, at any given time,
the wave function is of the form $\text{coeff.}[1/r-1/a(t)]+O(r)$ when the distance $r$ between
two fermions in opposite spin states is small. Here $a(t)$ is the instantaneous scattering length.
But since $mr_0^2/\hbar$ is much shorter than the other characteristic time scales of the system
[such as the Fermi time $m/(\hbar k_F^2)$], $t_\text{sweep}$ may still be comparable to or even
shorter than those other time scales, so that the whole quantum state may \emph{not} evolve adiabatically.

Given that the above instantaneous short-range boundary condition is satisfied, we can show
that the energy expectation value of the system changes at a rate
\begin{equation}\label{eq:dynamic}
\frac{\dif E}{\dif t}=\frac{\hbar^2\Omega C(t)}{4\pi m}\frac{\dif[-a(t)^{-1}]}{\dif t}
+\sum_\sigma\int\dif^3r\rho_{\sigma}(\vect rt)\frac{\partial V_\text{ext}(\vect rt)}{\partial t},
\end{equation}
where $C(t)$ is the instantaneous coefficient of the $1/k^4$ tail of the momentum distribution,
 $\rho_\sigma(\vect rt)$ is the spatial density of fermions in the spin state $\sigma$,
and for generality we also allow the external potential to change with time. We assume
that the inverse scattering length is a \emph{real} number at any instant of time.
We shall call Eq.~\eqref{eq:dynamic} \emph{dynamic sweep theorem}. This theorem
holds for any populations (no matter equal or unequal, small or large) of the two spin states,
any \emph{nonzero} scattering length, any quantum state, and any sweep rate, \emph{provided} that
the s-wave zero-range interaction model is still valid \cite{_zero_scattering_length}.
\begin{proof}
Let us first describe the time evolution of the quantum state. In the Schr\"{o}dinger picture,
a quantum state $\ket{\phi(t)}$ satisfies the Schr\"{o}dinger equation
\begin{equation*}
\dif\ket{\phi(t)}/\dif t=-(\I/\hbar)H(t)\ket{\phi(t)}
\end{equation*}
and the instantaneous short-range boundary condition
\begin{equation*}
\sum_{\vect k}\eta(\vect kt)c_{\vect q/2-\vect k\down}^{}c_{\vect q/2+\vect k\up}^{}\ket{\phi(t)}=0
~~~~(\text{any $\vect q$}),
\end{equation*}
where $H(t)=T+a(t)U+V(t)$ [$T$, $U$ and $V(t)$ are defined in Sec.~\ref{sec:proof}, except that now
$V_\text{ext}$ can change with time], and $\eta(\vect kt)\equiv\Lambda(\vect k)+L(\vect k)/4\pi a(t)$.

We then prove the equation for the conservation of probability:
\begin{equation*}
\frac{\dif\bracket{\phi(t)}{\phi(t)}}{\dif t}=0.
\end{equation*}
$\dif\bracket{\phi(t)}{\phi(t)}/\dif t=\brac{\phi(t)}\dif\ket{\phi(t)}/\dif t+\text{c.c.}=(-\I/\hbar)
\brac{\phi(t)}H(t)\ket{\phi(t)}+\text{c.c.}$, but $\brac{\phi(t)}H(t)\ket{\phi(t)}=\brac{\phi(t)}
[\eta(t)T+V(t)]\ket{\phi(t)}$ according to the energy theorem \cite{TanEnergyTheorem}, and it must be
real since the instantaneous scattering length is assumed to be real.
So $\dif\bracket{\phi(t)}{\phi(t)}/\dif t=0$.

Given the probability conservation, we may choose the normalized quantum state, such that
$\bracket{\phi(t)}{\phi(t)}\equiv1$. Then the system's energy expectation value is simply
\begin{equation*}
E(t)=\brac{\phi(t)}H(t)\ket{\phi(t)}.
\end{equation*}

Now consider two instants of time, $t$ and $t'$. For conciseness, we shall suppress the symbol
$t$ for a quantity at time $t$, and use a superscript $'$ for a quantity at $t'$ for distinction.
We then have
\begin{align*}
E'-E&=\brac{\phi'}H'\ket{\phi'}-\brac{\phi}H\ket{\phi}\\
&=\brac{\phi'}H'\ket{\phi'}^*-\brac{\phi}H\ket{\phi}\\
&=\brac{\phi}H'\ket{\phi'}^*-\brac{\phi'}H\ket{\phi}+R,
\end{align*}
where $R\equiv\Bigl\{\bigl(\brac{\phi'}-\brac{\phi}\bigr)H'\ket{\phi'}\Bigr\}^*
+\bigl(\brac{\phi'}-\brac{\phi}\bigr)H\ket{\phi}$,
\begin{align*}
\brac{\phi}H'\ket{\phi'}^*&=\brac{\phi}(T+a'U+V')\ket{\phi'}^*\\
&=\brac{\phi}(\eta T+a'\eta U+V')\ket{\phi'}^*\\
&=\brac{\phi}(\eta T+V')\ket{\phi'}^*\\
&=\brac{\phi'}(\eta T+V')\ket{\phi},
\end{align*}
and similary $\brac{\phi'}H\ket{\phi}=\brac{\phi'}(\eta' T+V)\ket{\phi}$. So
\begin{equation*}
E'-E=\brac{\phi'}\bigl[(\eta-\eta')T+(V'-V)\bigr]\ket{\phi}+R.
\end{equation*}

Finally we divide both sides of the above equation by $t'-t$, and take the limit $t'\rightarrow t$.
We get
\begin{equation*}
\frac{\dif E}{\dif t}=\brac{\phi}\biggl[-\frac{\dif\eta T}{\dif t}+\frac{\dif V}{\dif t}\biggr]
\ket{\phi}+\biggl[\frac{\dif\brac{\phi}}{\dif t}H\ket{\phi}+\text{c.c.}\biggr],
\end{equation*}
but $(\dif\brac{\phi}/\dif t)H\ket{\phi}=(\I/\hbar)(H\ket{\phi})^\dagger H\ket{\phi}
=(\I/\hbar)\bigl\lvert H\ket{\phi}\bigr\rvert^2$ is purely imaginary (or zero), so the two terms in the
last bracket cancel. Equation~\eqref{eq:dynamic} then easily follows
(note that the fermionic annihilation and creation operators are independent of time
in the Schr\"{o}dinger picture).

If the system is in a mixed state, it can be described by a density operator $\sum_i\rho_i
\ket{\phi_i(t)}\brac{\phi_i(t)}$, where the pure states $\ket{\phi_i(t)}$ obey the Schr\"{o}dinger
equation and the instantaneous short-range boundary condition, $\sum_i\rho_i=1$, and
$\bracket{\phi_i(t)}{\phi_j(t)}=\delta_{ij}$ (if this orthonormality condition is satisfied at
a time, it will be satisfied at all other time, because of the probability conservation).
The three terms in Eq.~\eqref{eq:dynamic} are all statistical averages of the
corresponding values in the pures states $\ket{\phi_i(t)}$ (with weights
$\rho_i$), so Eq.~\eqref{eq:dynamic} still holds.
\end{proof}

Note that in this proof, we \emph{never} write expressions like $H(t_1)\ket{\phi(t_2)}$
($t_1\neq t_2$),
since such entities contain delta function singularity, and the inner products between
them and any usual state vectors are divergent. It is only when the scattering length
in $H$ \emph{coincides} with the one in $\ket{\phi}$ that the delta function singularity
in $H\ket{\phi}$ vanishes, as is the case when $H$ and $\ket{\phi}$ are for the \emph{same}
instant of time.

Equation~\eqref{eq:dynamic} indicates that if $V_\text{ext}$ is held invariant,
but $-1/a(t)$ decreases, the expectation value of energy also decreases with time. This
appears to be incompatible with the common wisdom that if the sweep time is comparable to or
much shorter
than $ma^2/\hbar$, and $a$ is positive and decreases with time,
a fermionic dimer can be dissociated and the energy can increase.
The resolution to this paradox lies in the fact that in such a sweep, the dimer also has a
chance to evolve into a more tightly bound molecule, and the \emph{expectation
value} of the energy decreases. For more details, see Appendix \ref{app:A}.

Inspired by Appendix \ref{app:A}, we arrive at a general conjecture: if the scattering length
is suddenly changed from a large value $a_0$ to another large value $a_1$,
over a time scale $\Delta t$ ($\Delta t$ is much smaller
than $ma_1^2/\hbar$ and the other time scales of the system except $mr_0^2/\hbar$), 
then $\Omega C(t)$ is nearly a constant during this interval, and approximately equal to its value \emph{right before}
the scattering length begins to change. \emph{This is valid for any number of fermions}. We may exploit
this feature, to measure the integrated contact intensity as follows.

Consider any number of fermions in a trap having large scattering length $a_0$ and negligible $r_0$.
Suppose that the external trapping energy is $E_\text{ext}$, and the internal energy (kinetic plus interaction)
is $E_\text{int}$. If we suddenly change the scattering length to a large and \emph{negative}
(or infinite) value $a_1$,
and rapidly turn off the trapping potential at the same time, the gas will expand, and we can determine
its final energy $E'$ from the second moment of the spatial profile after a long enough time. The dynamic
sweep theorem, combined with the ansatz that $\Omega C(t)$ is nearly a constant in the \emph{short} interval
in which the scattering length changes, implies that
\begin{equation}\label{eq:sudden_sweep}
E'=E_\text{int}+\frac{\hbar^2\Omega C}{4\pi m}\biggl(\frac{1}{a_0}-\frac{1}{a_1}\biggr),
\end{equation}
where $\Omega C$ is the integrated contact intensity \emph{right before} the rapid change of the scattering length.
So if we plot $E'$ against $1/a_0-1/a_1$, we will get a \emph{straight line}, whose slope determines $\Omega C$,
and whose vertical intercept equals the initial internal energy of the gas.
It does not matter whether $a_0$ is positive, negative, or infinite. But $a_1$ had better be negative or infinite,
or some dimers will survive the expansion of the gas, complicating the measurement of energy.

Unlike another approach (depicted in Sec.~\ref{sec:intro}) to measure $\Omega C$, this one
makes use of the \emph{whole} spatial profile of the gas after its expansion,
and thus has a much lower noise-to-signal ratio.

If $\Delta t$ is very small, but not extremely so, it is possible to derive the lowest order correction
to Eq.~\eqref{eq:sudden_sweep}. Here we just show the result; the complete proof will appear later.
Suppose that
\begin{equation*}
b(t)\equiv 1/a(t)
\end{equation*}
approaches a constant $b_0$ sufficiently fast at
$t\rightarrow-\infty$, and approaches another constant $b_1$ sufficiently fast at $t\rightarrow+\infty$;
the time evolution of $b(t)$ occurs around $t=0$, over a characteristic time scale
$\Delta t$, which is small compared to the other time scales
of the system except $mr_0^2/\hbar$. If $mr_0^2/\hbar$ is negligible,
\begin{equation}
\Delta E=\frac{\hbar^2\Omega C_0}{4\pi m}\biggl(-\Delta b-\sqrt\frac{8\hbar}{\pi m}\iint\limits_{t'<t}
\!\sqrt{t-t'}\medspace\dot{b}(t')\dot{b}(t)\dif t'\dif t\biggr),
\end{equation}
where $\Delta E=E'-E_\text{int}$, $\Delta b=b_1-b_0$, $\dot{b}$ is the derivative of $b$ with respect to time,
and $C_0$ is the value that $C$ would have at $t=0$
\emph{if $b$ had remained equal to $b_0$ for all $t$}.
Note that this lowest order correction to Eq.~\eqref{eq:sudden_sweep} scales like $\sqrt{\Delta t}$,
and is independent of the state of the fermions; the next order correction
will most likely scale like $\Delta t$.
In the derivation of this formula, one exploits the knowledge that $C(t)$ only changes a little
during the fast ramp of the inverse scattering length. Note also that this correction term to $\Delta E$ is not
always negative; if, for example, $b$ suddenly increases from $b_0$ to $b_2$, and after a short period $\Delta t$
it suddenly decreases from $b_2$ to $b_1$, then the correction term is positive.

The author thanks T.~L.~Ho for a stimulating suggestion.
The author thanks K.~Levin for the introduction to ultracold Fermi gases,
and S.~Giorgini for communication. 

\appendix
\section{\label{app:A}Dissociation of a fermionic dimer after a sudden change of the scattering length}
Our purpose here is to illustrate the dynamic sweep theorem with a simple two-body problem.

A dimer with fermionic scattering length $a_0>0$ (and $a_0\gg r_0$) is initially
at rest. The initial value of $\Omega C$ is easily calculated to be
$8\pi/a_0$. At time $t\approx0$, the scattering length is
suddenly changed to $a$ ($\lvert a\rvert\gg r_0$), over a time scale that is both much shorter
than $\min(ma_0^2/\hbar, ma^2/\hbar)$ and much longer than $mr_0^2/\hbar$. What is the energy
expectation value of the system afterwards? Suppose $V_\text{ext}\equiv0$.

Answer: since $C(t)$ can only change continuously with time (we will verify this below), it can
be regarded as almost a constant during the short interval during which the inverse scattering length
is changing. Equation~\eqref{eq:dynamic} can then be integrated to yield $\Delta E=\hbar^2\Omega C(0)
(1/a_0-1/a)/(4\pi m)=(\hbar^2/m)2b_0(b_0-b)$, where
\begin{equation*}
b_0\equiv1/a_0,~~~~b\equiv1/a.
\end{equation*}
The system's initial energy is $-(\hbar^2/m)b_0^2$. So its energy expectation value after
the change of the scattering length should be
\begin{equation}
E=(\hbar^2/m)(b_0^2-2b_0b)~~~~~~(t>0).
\end{equation}

Now we confirm this result with the concrete calculation. The system's wave function is
\begin{equation}
\psi(\vect r,t)\equiv\frac{\varphi(r,t)}{\sqrt{2\pi a_0}~r}
\end{equation}
which is normalized: $\int\dif^3r\lvert\psi(\vect r,t)\rvert^2=1$,
and $\varphi(r,t)=\exp(-b_0r+\I\hbar b_0^2t/m)$ for $t<0$. Here $r$ is the distance
between the two constituent fermions.

For $t>0$, we can expand the wave function in terms of the new energy eigenstates
(associated with the scattering length $a$), and get
\begin{multline}\label{eq_ap_a:1}
\varphi(r,t)=A\exp(-br+\I\hbar b^2 t/m)\\+\int_{0}^\infty B(k)\sin[kr-\alpha(k)]\exp(-\I\hbar k^2t/m)\dif k,
\end{multline}
where the scattering phase shift
$\alpha(k)=\arccos\frac{b/k}{\sqrt{1+b^2/k^2}}$ is in the range $(0,\pi)$;
this ensures that the phase shift does not jump when we pass through the unitarity limit.
The base functions satisfy orthogonality conditions
$\int_0^\infty\exp(-br)\sin[kr-\alpha(k)]\dif r=0$ (if $a>0$) and
$\int_0^\infty\sin[kr-\alpha(k)]\sin[k'r-\alpha(k')]\dif r=(\pi/2)\delta(k-k')$ (any $a$),
and $k,k'>0$.

The coefficients
$A$ and $B(k)$ can be determined from the projection of $\varphi(r,0^-)$ onto the new energy eigenstates,
because the wave function can only evolve continuously, $\varphi(r,0^-)=\varphi(r,0^+)$. The result is
\begin{align}
A&=\frac{2b\theta(b)}{b+b_0},\label{eq_ap_a:2}\\
B(k)&=\frac{2(b-b_0)k}{\pi(k^2+b_0^2)\sqrt{k^2+b^2}}\label{eq_ap_a:3},
\end{align}
where $\theta(b)=1$ if $b\geq0$ and $\theta(b)=0$ if $b<0$.

These functions satisfy the probability conservation condition
$P_0+P_1=1$,
where \begin{equation*}P_0=A^2b_0/b=4bb_0\theta(b)/(b+b_0)^2\end{equation*}
is the probability for the dimer to remain in the bound
state, and $P_1=\int\dif P_1=\int_0^\infty\pi b_0[B(k)]^2\dif k=(b^2+b_0^2-2\lvert b\rvert b_0)/(b+b_0)^2$
is the probability for it to be dissociated.

The system's energy expectation value for $t>0$ is clearly
\begin{equation}
E=(\hbar^2/m)\biggl\{-b^2P_0+\int_0^\infty k^2\cdot\pi b_0[B(k)]^2\dif k\biggr\}.
\end{equation}
Substituting $P_0$ and $B(k)$ into this expression, we find $E=(\hbar^2/m)b_0(b_0-2b)$,
consistent with our previous expectation based on the dynamic sweep theorem.

Now we verify the continuity of the contact intensity. 

For any given $t<0$, $\varphi(r,t)=\exp(\I\hbar b_0^2t/m)(1-b_0r)+O(r)$ when $r\rightarrow0$.

For any given $t>0$, according to Eqs.~\eqref{eq_ap_a:1},
\eqref{eq_ap_a:2} and \eqref{eq_ap_a:3}, $\varphi(r,t)=\varphi(0,t)(1-br)+O(r)$ when $r\rightarrow0$, where
\begin{widetext}\begin{equation}
\varphi(0,t)=\frac{b_0\exp(\I b_0^2\hbar t/m)\biggl\{1-\erf\Bigl[\exp(\I\pi/4)b_0\sqrt{\hbar t/m}\Bigr]\biggr\}
+b\exp(\I b^2\hbar t/m)\biggl\{1+\erf\Bigl[\exp(\I\pi/4)b\sqrt{\hbar t/m}\Bigr]\biggr\} }
{b_0+b}~~~~(t>0).
\end{equation}\end{widetext}
$\erf(z)\equiv\bigl(2/\sqrt{\pi}\bigr)\int_0^z\exp(-w^2)\dif w$ is the error function. The value of $\varphi(0,t)$
($t>0$) in the special case $b=-b_0$ is given by the limit $b\rightarrow-b_0$.
In any case, $\lim_{t\rightarrow 0^-}\varphi(0,t)=\lim_{t\rightarrow 0^+}\varphi(0,t)=1$, and this means
that $\Omega C(t)=(8\pi/a_0)\lvert\varphi(0,t)\rvert^2$
is continuous. But note that $[(\partial/\partial r)\varphi(r,t)]_{r=0}$ is
discontinuous at $t=0$.

The Fourier transform of $\varphi(0,t)$ ($-\infty<t<\infty$) is simple:
\begin{equation}
\widetilde{\varphi}(0,\omega)=\frac{p_0^{}-p}{\bigl(\omega-\I\epsilon+p_0^2\bigr)
\bigl(\sqrt{\omega+\I\epsilon}-\I p\bigr)},
\end{equation}
where $p_0^{}\equiv\sqrt{\hbar/m}/a_0$, $p\equiv\sqrt{\hbar/m}/a\neq p_0^{}$, and $\epsilon\rightarrow0^+$.

This two-body problem also helps to assure us that the momentum distribution
of fermions with large scattering length ($\lvert a_0\rvert\gg r_0$) is indeed \emph{physically measurable}
\cite{Hammer_dispute}.
According to Eq.~\eqref{eq_ap_a:3}, \emph{if the dimer is dissociated}, the momentum distribution of the
fermions in the limit $t\rightarrow+\infty$ (corresponding to a directly-measurable
spatial probability distribution) is
\begin{equation}
n_{\vect k\sigma}\propto\frac{1}{(1+a^2k^2)(k^2+1/a_0^2)^2},
\end{equation}
which in the limit $\lvert a\rvert\ll a_0$ approaches the momentum distribution of the fermions
in the original dimer.
\bibliography{adiabatic}

\begin{thebibliography}{43}
\bibitem[{_an()}]{_anyT}
\bibinfo{note}{The temperature should satisfy
  $T_\text{temp}\ll\hbar^2/(k_B^{}mr_0^2)$, or the s-wave contact interaction
  model breaks down (the subscript is used to distiguish temperature from the
  kinetic energy operator $T$). For $^{40}$K atoms near a wide Feshbach
  resonance, $r_0$ is of the order a few nanometers \cite{Gribakin}, and
  $\hbar^2/(k_B^{}mr_0^2)\sim 1\mathrm{mK}$.}
\bibitem[{_l_()}]{_l_updown}
\bibinfo{note}{Usually $l_{\up\down}$ is the average interparticle spacing
  $1/k_F^{}$; but if there are tightly bound fermionic dimers ($1/k_F^{}\gg
  a\gg r_0$), $l_{\up\down}\sim a$.}
\bibitem[{_tu()}]{_tune}
\bibinfo{note}{As is well known, for ultracold atoms near a Feshbach resonance
  controlled by a magnetic field $B$, when $B$ is tuned, $1/a$ changes.}
\bibitem[{_Fe()}]{_FermiPseudopotential}
\bibinfo{note}{Although such a pseudopotential was previously used
  \cite{HuangYang, LeeHuangYang} in the context of hard sphere interaction, it
  is a much better approximation for particles interacting with a large
  scattering length and a small interaction radius; in the limit of zero
  interaction radius (but a nonzero scattering length), this pseudopotential
  becomes exact.}
\bibitem[{_BE()}]{_BEC_caveat}
\bibinfo{note}{A practical caveat is that in the BEC limit, the many-body
  corrections to $C$ [Eq.~\eqref{eq:C_BEC}] are small and difficult to detect
  in experiment.}
\bibitem[{_co()}]{_contact}
\bibinfo{note}{There are of course errors introduced by the omission of the
  deviations (due to a small but nonzero $r_0$) from the s-wave contact
  interaction, but well-controlled corrections should be possible.}
\bibitem[{_en()}]{_energy_change}
\bibinfo{note}{To change the system's scattering length, there must be some
  physical means. For ultracold atoms near a Feshbach resonance, we can, as is
  well known, achieve this by tuning the external magnetic field $B$. The
  system will absorb energy from the electromagnetic field or release energy to
  it when $B$ is tuned, and the law of energy conservation will not be
  violated.}
\bibitem[{_mr()}]{_mr02_example}
\bibinfo{note}{It is instructive to estimate this time scale for an actual
  physical system. For example, for $^{40}$K atoms, $r_0$ is of the order a few
  nanometers \cite{Gribakin}, and $mr_0^2/\hbar$ is of the order $1\sim10$
  nanoseconds.}
\bibitem[{_ze()}]{_zero_scattering_length}
\bibinfo{note}{If the scattering length is tuned all the way to zero, this
  theorem will become inapplicable at some point, when the condition $\lvert
  a\rvert\gg r_0$ is broken.}
\end{thebibliography}
\end{document}